\documentclass[10pt,aps,prb,twocolumn,longbibliography,superscriptaddress,floatfix]{revtex4-2}
\usepackage{comment}
\usepackage{graphicx} 
\usepackage{xcolor}
\usepackage[colorlinks=true,linkcolor=blue,citecolor=blue,urlcolor=blue]{hyperref}
\usepackage[normalem]{ulem} 

\begin{document}
\title{Complex Magnetic Behavior in RuO$_2$ Thin Films Driven by Strain and Substrate Effects}
\author{Mojtaba Alaei}
\email{m.alaei@skoltech.ru}
\affiliation{Materials Discovery Laboratory, Skolkovo Institute of Science and Technology, Bolshoy Boulevard 30, Building 1, Moscow 121205, Russia}
\affiliation{Department of Physics, Isfahan University of Technology, Isfahan 84156-83111, Iran}
\author{Nafise Rezaei}
\affiliation{Materials Discovery Laboratory, Skolkovo Institute of Science and Technology, Bolshoy Boulevard 30, Building 1, Moscow 121205, Russia}
\author{Ilia Mikhailov}
\affiliation{Materials Discovery Laboratory, Skolkovo Institute of Science and Technology, Bolshoy Boulevard 30, Building 1, Moscow 121205, Russia}
\author{Artem R. Oganov}
\affiliation{Materials Discovery Laboratory, Skolkovo Institute of Science and Technology, Bolshoy Boulevard 30, Building 1, Moscow 121205, Russia}
\author{Alireza Qaiumzadeh}
\email{alireza.qaiumzadeh@ntnu.no}
\affiliation{Center for Quantum Spintronics, Department of Physics, Norwegian University of Science and Technology, NO-7491 Trondheim, Norway}

\date{\today}

\begin{abstract}
Ruthenium dioxide (RuO$_2$) has been proposed as a prototypical metallic $d$-wave altermagnet, a Néel-ordered compensated antiferromagnetic state exhibiting nonrelativistic momentum-dependent spin splitting; yet, its magnetic ground state remains controversial both theoretically and experimentally. Using comprehensive first-principles calculations, we investigate RuO$_2$ thin films with (110), (100), and (001) orientations, both (un)strained freestanding and supported on a TiO$_2$ substrate. We show that emergent magnetic moments in RuO$_2$ thin films are highly fragile, strongly influenced by strain, surface orientation, and atomic relaxation, while also being highly sensitive to the choice of the Brillouin-zone integration scheme. 
We find that none of the thin film structures considered can stabilize a compensated antiferromagnetic order; therefore, an altermagnetic ground state cannot be realized. Instead, substrate-supported RuO$_2$ films on TiO$_2$ exhibit pronounced layer- and site-dependent magnetic moment variations and incomplete compensation between the two antiferromagnetically coupled Ru moments, yielding a \emph{ferrimagnetic-like} behavior. On the other hand, freestanding RuO$_2$ films display complex magnetic structures depending on their orientation and applied strain, with distinct behavior at the surfaces and in the inner layers.   
Our results reconcile conflicting theoretical and experimental reports and underscore the sensitivity of RuO$_2$ magnetism to structural and methodological details. 
\end{abstract}

\maketitle

\section{Introduction}
Finding and reliably identifying the true magnetic ground state of quantum materials remains one of the central challenges in condensed-matter physics and materials science \cite{tellez2024systematic,aldea2025challenges,PhysRevB.111.104416}. This task is particularly demanding in itinerant transition-metal oxides, where competing exchange interactions, electron itinerancy, and structural distortions often place the system close to multiple nearly degenerate magnetic instabilities \cite{goodenough2014perspective,solovyev2003}. Even subtle perturbations, such as epitaxial strain, reduced dimensionality, surface termination, or details of electronic-structure methodology, can tip the balance between nonmagnetic, ferro-, ferri-, and antiferromagnetic states. Establishing the genuine magnetic ground state is, therefore, not only a fundamental problem but also a prerequisite for designing functional materials for spin-based technologies, where transport, symmetry, and magnetic order are intimately intertwined.

Recently, unconventional antiferromagnetic (AFM) materials exhibiting nonrelativistic, momentum-dependent spin splitting have been identified as a symmetry-distinct class of antiferromagnets beyond conventional spin-degenerate AFM systems \cite{noda2016momentum,hayami2019,naka2019spin,PhysRevB.102.014422,Sinova-3,Sinova-1,Sinova-2, MnTe_nature, PhysRevB.111.104416}. In these systems, the spin splitting originates not from the relativistic spin–orbit interaction but from the spin space/point group symmetry, which allows for exchange-driven lifting of spin degeneracy while preserving zero net magnetization. The resulting momentum-dependent spin splitting is constrained by crystal symmetry and can be classified according to its angular-momentum character in the electronic band structure. In particular, unconventional collinear AFM materials with even-parity spin splitting of $d$-, $g$-, or $i$-wave symmetry, collectively termed altermagnets, retain fully compensated magnetic moments in real space while exhibiting sign-changing spin polarization across the Brillouin zone \cite{Sinova-1,Sinova-2}. This symmetry-enforced structure of the spin splitting distinguishes it from conventional $s$-wave AFMs, in which combined inversion or parity and time-reversal symmetries protect spin degeneracy throughout momentum space.

Among these unconventional collinear AFMs, the $d$-wave class is particularly attractive because its symmetry allows for directional spin-polarized transport via the spin-splitter effect and spin splitting torque \cite{Sinova-4,zwz9-l7wf}. Despite rapid progress in identifying candidate materials, many proposed altermagnets are insulating, which limits their immediate applicability in spin-transport devices. For practical applications, metallic or semiconducting systems are considerably more desirable, as they enable direct electrical control and detection of spin-polarized currents. 
RuO$_2$ has been proposed as a prominent metallic candidate, with symmetry analysis predicting $d$-wave altermagnetic order. Early \emph{ab initio} calculations \cite{PhysRevLett.118.077201, PhysRevB.99.184432,RuO2_PRL,RuO2_advance} and a few direct experimental observations and indirect quantum transport measurements  \cite{PhysRevLett.118.077201, PhysRevLett.122.017202, RuO2_AHE1,RuO2_advance,RuO2_AHE2,He2025-3,Ono2024-5,Zhang2025-7,jeong2025-11,Jeong2025-13,howzen-19} support this possibility. However, more recent high-resolution measurements, such as X-ray linear dichroism, nuclear magnetic resonance, spin-ARPES, and neutron diffraction; along with advanced first-principles studies, have challenged this picture, indicating that bulk/thin-film RuO$_2$ may, in fact, be nonmagnetic or, at most, extremely close to a magnetic \cite{Mazin2024-6,Plouff2025-2,Sato2026-10,sym-break2025,vqrj-21}. This apparent contradiction has fueled an ongoing debate regarding the true magnetic ground state of RuO$_2$ in bulk and thin films \cite{choi2026exploring,jeong2026,RuO2_nomag1,RuO2_nomag2,RuO2_nomag3, RuO2_nomag4, RuO2_nomag5, RuO2_nomag6,Lu2026-9,Akashdeep-16,Chen-17,Lin-20,lee2026}.

\emph{Ab initio} methods based on density functional theory (DFT) are powerful tools for determining magnetic ground states. Standard DFT calculations without an on-site Hubbard $U$ correction predict a nonmagnetic ground state for bulk RuO$_2$, whereas DFT+$U$ calculations with $U$ values of $U > 0.6$ eV stabilize a metallic altermagnetic state~\cite{RuO2_nomag3, Fragile_2025,Mazin2024-6}. Although the applicability of DFT+$U$, which effectively enhances the localization of Ru 4$d$ electrons, may be questioned for a system with predominantly itinerant character, the fact that magnetism emerges already for relatively small $U$ values indicates that RuO$_2$ lies in close proximity to a magnetic instability. This pronounced sensitivity implies that its electronic and magnetic properties can be strongly influenced by external perturbations, including strain, symmetry breaking at surfaces and interfaces, and chemical doping~\cite{Fragile_2025,Strain_Engineering,Strain1,sym-break2025,RuO2_110-mag1}.

Recent experimental measurements have focused on the epitaxial growth of RuO$_2$ thin films on TiO$_2$ substrates, highlighting the importance of reduced dimensionality and substrate-induced effects on their magnetic properties. These developments underscore the need for a clear microscopic understanding of magnetism in RuO$_2$ from first principles. 
Although several recent \emph{ab initio} investigations have focused on the magnetism of RuO$_2$ thin films~\cite{sym-break2025, magnetization_001}, significant gaps remain in our understanding of the detailed magnetic behavior of these systems.

In this study, we perform comprehensive DFT calculations to achieve a microscopic understanding of the magnetic behavior of both freestanding RuO$_2$ slabs and epitaxial RuO$_2$ thin films on TiO$_2$ with (110), (001), and (100) orientations. 
Since the (110) orientation is the most thermodynamically stable surface for both TiO$_2$ and RuO$_2$~\cite{TiO2_110_stable,RuO2_110_stable}, our analysis places particular emphasis on this configuration. For this orientation, we systematically assess the influence of Brillouin-zone sampling, structural relaxation, induced magnetic moments, and competing magnetic ordering patterns in order to obtain a robust and quantitatively reliable determination of the magnetic ground state.

Using systematically converged $k$-point sampling, we show that epitaxial RuO$_2$ on a TiO$_2$ substrate fails to stabilize a compensated altermagnetic state in all three orientations. Instead, it adopts a ferrimagnetic-like ground state characterized by a finite net magnetization and pronounced layer-dependent variations in the magnetic moment.
In contrast, freestanding thin films remain nonmagnetic in the absence of strain, with only small magnetic moments emerging in a few surface-near layers; notably, the (001) surface remains entirely nonmagnetic. 
Under applied strain, freestanding films develop weak magnetic moments with varying amplitudes across the slab. Nevertheless, for the (001) orientation, the inner layers remain nonmagnetic, and only small surface-localized moments appear under strain.

\section{Model Structures}
\begin{figure}
    \centering
    \includegraphics[width=1.0\linewidth]{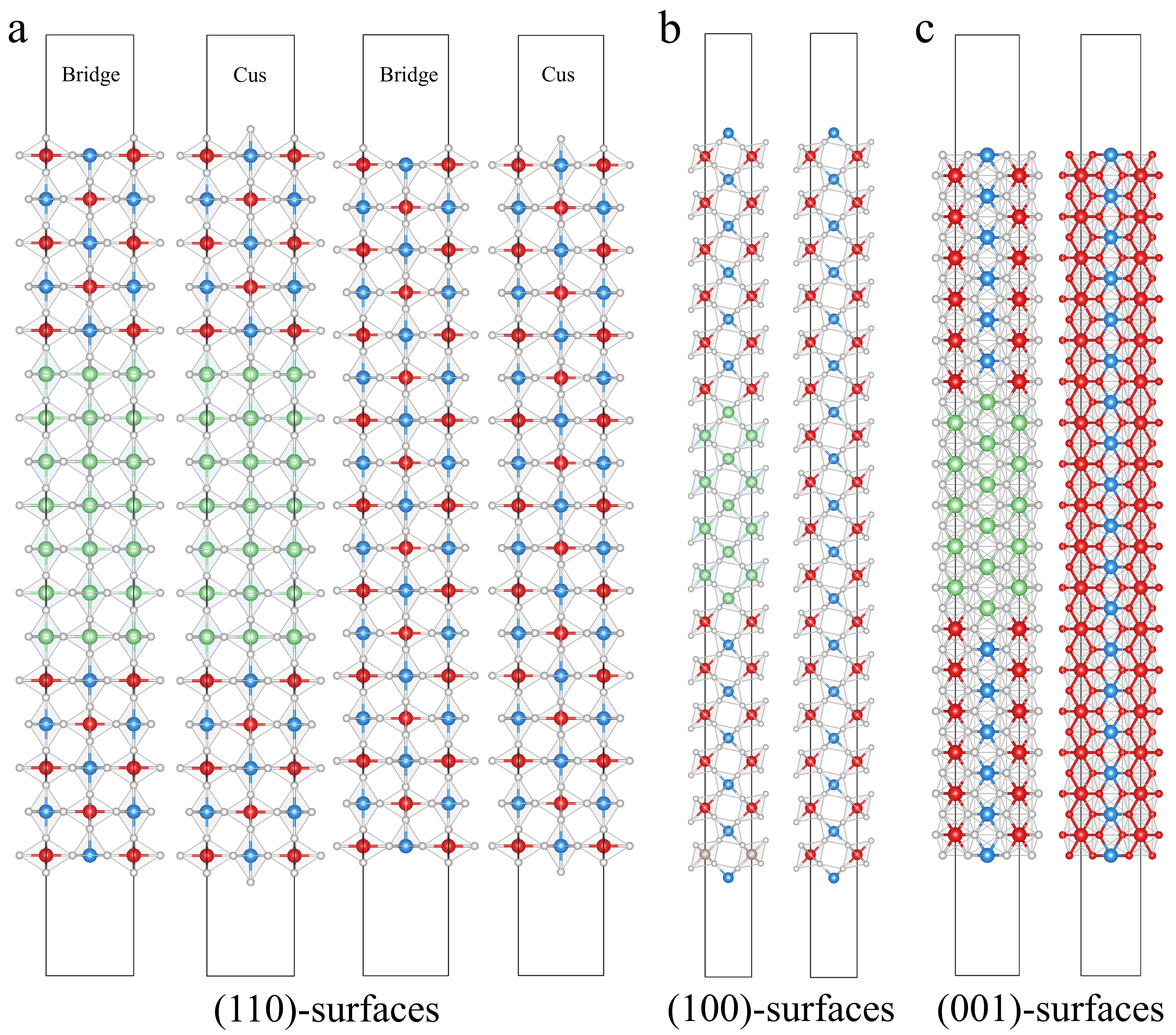}
    \caption{Different surface orientations of RuO$_2$ thin films considered in this work, for both freestanding and substrate-supported configurations. For the RuO$_2$(110)/TiO$_2$(110) heterostructure, several film thicknesses were examined; the 5-L configuration is shown as a representative example. Blue and red spheres denote Ru atoms carrying different magnetic moments in the assumed AFM configurations discussed in the main text (labelled Ru$_1$ and Ru$_2$). Ti and O atoms are shown in green and white, respectively.}
    \label{fig:struct}
\end{figure}

We investigate three RuO$_2$ thin-film surface orientations, (110), (100), and (001); considered both as freestanding slabs and as substrate-supported films grown on TiO$_2$ (see Fig.~\ref{fig:struct}). The RuO$_2$/TiO$_2$ heterostructures are constructed using the experimental crystal structure of TiO$_2$. For the RuO$_2$(110)/TiO$_2$(110) system, recently studied experimentally, we examine RuO$_2$ film thicknesses ranging from one to seven Ru atomic layers.
For the freestanding RuO$_2$(110) thin film, a 17-layer Ru slab is employed to ensure convergence with respect to slab thickness and to minimize spurious interactions between the two surfaces. 
For RuO$_2$(100)/TiO$_2$(100) and RuO$_2$(001)/TiO$_2$(001) heterostructures, the RuO$_2$ film thickness is fixed at 12 Ru atomic layers. In the case of freestanding RuO$_2$(100) and RuO$_2$(001) thin films, thicker slabs consisting of 33 and 35 Ru atomic layers, respectively, are used to achieve well-converged surface properties.

The RuO$_2$/TiO$_2$ heterostructure slabs were constructed by placing RuO$_2$ layers symmetrically on both sides of the TiO$_2$ substrate. This symmetric geometry avoids artificial effects associated with asymmetric slab models, where depositing RuO$_2$ on only one side can introduce a net dipole moment across the slab. In such cases, spin-polarized calculations may yield spurious magnetic moments on the Ti atoms at the exposed surface. By eliminating the net dipole moment, the symmetric slab geometry suppresses these artifacts and ensures a physically meaningful description of the magnetic properties.

For the RuO$_2$(110) thin film, three distinct $(1 \times 1)$ terminations of the rutile structure have been reported in the literature~\cite{RuO2_110_Karsten}. In this work, we consider two of these terminations; see Fig. \ref{fig:struct}a.
The relative stability of these terminations depends on the oxygen chemical potential, reflecting the oxygen availability during growth or operation~\cite{RuO2_110_Karsten}. Under oxygen-rich conditions, the stoichiometric RuO$_2$(110)–O$^{\mathrm{bridge}}$ termination is the most stable and is widely regarded as the preferred configuration of the rutile (110) surface. Under oxygen-poor conditions, the surface instead favors the RuO$_2$(110)–O$^{\mathrm{cus}}$ termination, characterized by coordinatively unsaturated Ru sites (cus). The third possible termination, RuO$_2$(110)–Ru, is not considered here, as it is unstable even under strongly oxygen-deficient conditions.

Since our explicit comparison of oxygen-rich and oxygen-poor terminations for the (110) orientation reveals no significant changes in the magnetic ordering, we limit our analysis of the (100) and (001) surfaces to terminations without oxygen atoms atop the Ru sites; see Figs. \ref{fig:struct}b and Fig. \ref{fig:struct}c.

The RuO$_2$(110) orientation displays a repeating three-layer stacking sequence consisting of two oxygen-only layers and one mixed Ru–O layer. A similar three-layer periodicity is found for the RuO$_2$(100) orientation, composed of two oxygen-only layers and one ruthenium-only layer. In contrast, for the (001) orientation, each atomic layer contains both Ru and O atoms.
To adopt a unified notation for slab thickness, we define the thickness solely by the number of Ru-containing layers. For example, for the (110) surface, with its sequence of two O layers followed by one mixed Ru–O layer, we use the notation 1-L, 2-L, \ldots, $n$-L instead of the conventional tri-layer notation (1-TL, 2-TL, \ldots, $n$-TL).

\section{Results}
\subsection{Magnetism and k-point sampling sensitivity}
A recent \emph{ab initio} study~\cite{Fragile_2025} investigated the effect of strain on the magnetic properties of bulk RuO$_2$, showing that lattice expansion stabilizes magnetic order, an effect also observed in related systems~\cite{OsO2_2025}. Moreover, the authors reported that the calculated magnetic moment is highly sensitive to the choice of \textit{k}-point mesh, displaying pronounced fluctuations. This sensitivity was attributed to the fragile and unconventional character of magnetism in RuO$_2$.

Here, we demonstrate that these fluctuations largely originate from the specific Brillouin-zone integration scheme employed. To substantiate this conclusion, we evaluate the  total absolute magnetization of RuO$_2$ under an effectively tensile-strained configuration obtained by adopting the lattice parameters and atomic positions of rutile TiO$_2$. 
Here, the total absolute magnetization is defined as the sum over the absolute magnetic moments of all atoms in the unit cell, $|M|= \sum^{N_{\rm{atoms}}}_{i=1} |M^i|= \int_\mathrm{cell} \left| n_\mathrm{up}(\mathbf{r}) - n_\mathrm{down}(\mathbf{r}) \right| \, d^3r$,
where $n_\mathrm{up}(\mathbf{r})$ and $n_\mathrm{down}(\mathbf{r})$ are the spin-up and spin-down charge densities.
In this regime, RuO$_2$ stabilizes a compensated AFM order and consequently exhibits altermagnetic behavior.

We tested several Brillouin-zone integration schemes across different \textit{k}-point meshes, including smearing-based approaches such as Marzari–Vanderbilt–DeVita–Payne cold smearing~\cite{cold} and Fermi–Dirac smearing. Because smearing methods depend on the choice of the broadening parameter ($\sigma$), we also employed the optimized tetrahedron method~\cite{opt_tet}, which enables accurate Brillouin-zone integration without smearing. The resulting total absolute magnetizations were then systematically compared across all integration schemes and \textit{k}-point grids.

\begin{figure}
    \centering
    \includegraphics[width=1.0\linewidth]{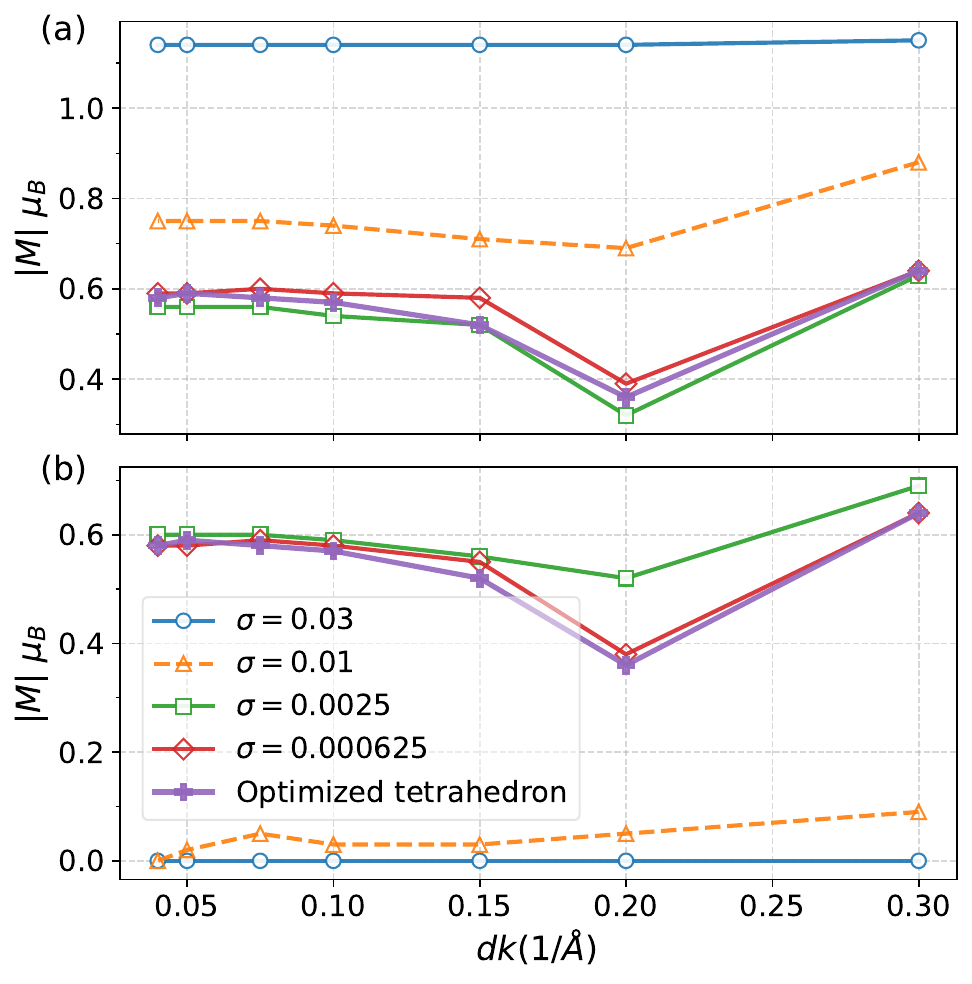}
\caption{Sensitivity of the total absolute magnetization (and, consequently, the Ru magnetic moments) to \textit{k}-point sampling in bulk rutile RuO$_2$ under tensile strain. The calculations are performed for bulk RuO$_2$ constructed using the lattice parameters and Wyckoff positions of rutile TiO$_2$, which leads to a tensile strain and stabilizes an AFM state. \textit{k}-point densities range from 0.05~Å$^{-1}$ (corresponding to a $27 \times 27 \times 42$ Monkhorst–Pack mesh) to 0.30~Å$^{-1}$ (corresponding to a $5 \times 5 \times 8$ mesh).
(a) The total absolute magnetization as a function of the Monkhorst–Pack \textit{k}-point grid calculated using Marzari–Vanderbilt–DeVita–Payne cold smearing for different smearing widths $\sigma$ (in Rydberg). Results obtained with the optimized tetrahedron method, which does not require a smearing parameter, are shown for comparison.
(b) Same analysis performed using Fermi–Dirac smearing with a smearing amplitude $\sigma$.}
    \label{fig:m_vs_k}
\end{figure}

As shown in Fig.~\ref{fig:m_vs_k}, the calculated total absolute magnetization (and, consequently, the magnetic moments of the Ru atoms) is highly sensitive to both the choice of the smearing scheme and the value of the broadening parameter $\sigma$. For example, Fermi–Dirac smearing with $\sigma = 0.01$–$0.03$~Ry yields nearly vanishing magnetic moments, whereas cold smearing under the same conditions produces a large total absolute magnetization of approximately 1.0~$\mu_\mathrm{B}$. These results underscore the importance of carefully selecting the Brillouin-zone integration scheme and broadening parameter when investigating magnetism in RuO$_2$. 

In subsequent calculations, we performed structural relaxations using a small electronic broadening ($\sigma = 0.00125$~Ry), which yields results nearly identical to those obtained with the optimized tetrahedron method. Final total-energy and magnetic-property calculations were then carried out using the optimized tetrahedron method to ensure an accurate treatment of Brillouin-zone integration.

\subsection{RuO$_2$(110)/TiO$_2$(110) heterostructure}
Recent experimental investigations of the electronic band structure and transport properties of RuO$_2$ have focused on RuO$_2$(110) thin films grown on TiO$_2$(110), owing to their favorable epitaxial matching and structural stability.
We first, therefore, investigate RuO$_2$(110) thin films with thicknesses ranging from one to seven Ru-containing layers grown on a TiO$_2$(110) substrate. Site-resolved magnetic moments are obtained by Bader partitioning of the spin density. In the AFM phase of bulk RuO$_2$, the rutile structure contains two inequivalent Ru sites that form distinct magnetic sublattices, which we label Ru$_1$ and Ru$_2$. These sites are indicated in Fig.~1 by blue (Ru$_1$) and red (Ru$_2$) spheres.

\begin{figure}
    \centering
    \includegraphics[width=1.0\linewidth]{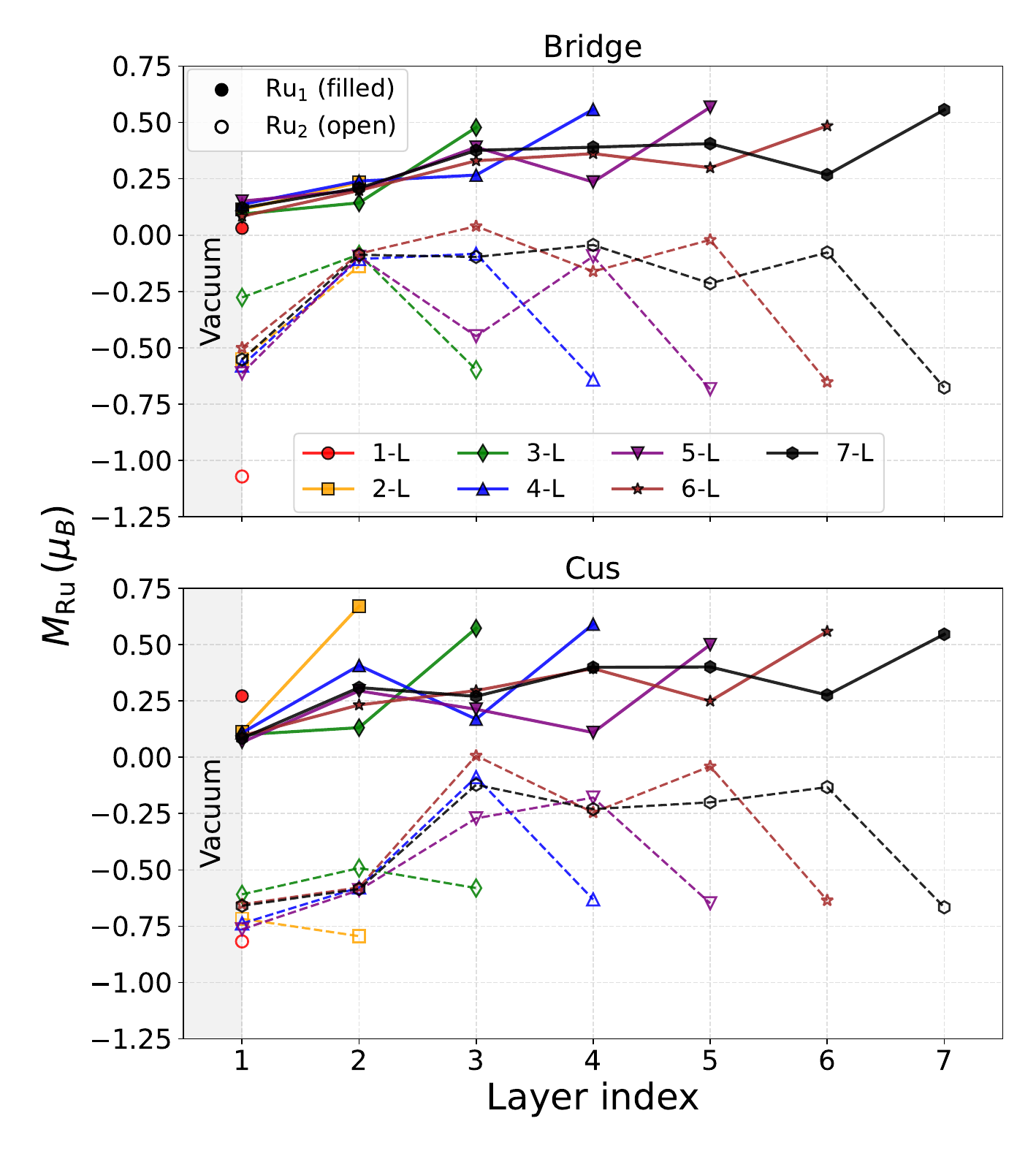}
    \caption{Layer-resolved magnetic moments of Ru$_1$ and Ru$_2$ for film thicknesses between one and seven layers, for Bridge (top) and Cus (bottom) terminations. Filled and open symbols denote Ru$_1$ and Ru$_2$, respectively, for RuO$_2$(110) thin films on a TiO$_2$(110) substrate. Data corresponding to the same film thickness are connected by lines of the same color to guide the eye. Opposite magnetic moments of Ru$_1$ and Ru$_2$ atoms do not cancel each other and thus system behave like an ferrimagnet state, with varying magnetic moments at each layer, rather than a compensated AFM order in altermagnets.}
    \label{fig:m_ru110}
\end{figure}

Figure~\ref{fig:m_ru110} presents the magnetic moments of Ru$_1$ and Ru$_2$ at each layer as a function of the film thickness for Bridge (top panel) and Cus (bottom panel) terminations. The opposite magnetic moments on Ru$_1$ and Ru$_2$, in both cases, do not completely cancel. 
Consequently, the RuO$_2$(110) thin film on a TiO$_2$(110) substrate exhibits a finite net magnetization, consistent with ferrimagnetic-like behavior rather than the compensated AFM order characteristic of altermagnets.
We call it ferrimagnetic-like state since, contrary to a ferrimagnetic state, the magnetic moments exhibit layer-dependent amplitudes: the moments on both Ru$_1$ and Ru$_2$ in different layers oscillate as the thickness increases, although without a strictly periodic pattern. These variations are most pronounced at the outermost surface layer and at the RuO$_2$/TiO$_2$(110) interface, underscoring the strong influence of surface and interface effects on the magnetic properties.

To further characterize the magnetic behavior of RuO$_2$(110)/TiO$_2$(110), we examine three key factors: the influence of the initial magnetic configuration, the role of atomic relaxation, and the impact of induced magnetic moments.

Because standard DFT calculations cannot systematically sample the full magnetic energy landscape, the choice of the initial magnetic configuration is particularly important for RuO$_2$, where magnetic moments, if any, are intrinsically small. Exhaustive exploration of all possible magnetic initializations for different thicknesses is computationally prohibitive; we therefore perform a detailed analysis for a representative slab with a thickness of 3-L.

For (110) orientation, each layer contains two inequivalent Ru atoms (Ru$_1$ and Ru$_2$), yielding a total of six magnetic sites and ten distinct AFM configurations. We systematically examined all these AFM arrangements. Each configuration was initially imposed using magnetic constraints to stabilize the desired spin pattern; the constraints were subsequently released to allow full self-consistent relaxation. Among all tested configurations, the one initialized with antiparallel magnetic moments between Ru$_1$ and Ru$_2$ within each layer, i.e., a checkerboard configuration, consistently relaxes to the lowest-energy state, characterized by opposite signs of the magnetic moments on Ru$_1$ and Ru$_2$, with incomplete compensation. The remaining AFM initializations either converge to higher-energy solutions or evolve into the same checkerboard configuration upon relaxation. Moreover, for several slab thicknesses, calculations starting from a ferromagnetic state also converge to this checkerboard-type AFM order. These results indicate that the ground state generally favors antiparallel alignment between Ru$_1$ and Ru$_2$ within each layer; accordingly, this antiparallel arrangement is adopted as the initial magnetic configuration in all subsequent calculations.

The only deviation from this trend occurs in the 6-L slab, where the Ru atoms in the third layer develop weak magnetic moments of the same sign on the two sublattices, despite being initialized in an AFM configuration. This behavior could be associated with small structural relaxation effects. In the third layer, the magnetic moment on Ru$_2$ is significantly smaller than that on Ru$_1$. The magnetic moments in the third layer are 0.33~$\mu\mathrm{B}$ on Ru$_1$ and 0.04~$\mu\mathrm{B}$ on Ru$_2$ for the Bridge configuration, whereas for the Cus configuration, the corresponding values are 0.30~$\mu\mathrm{B}$ and 0.007~$\mu\mathrm{B}$, respectively.

\begin{figure}
    \centering
    \includegraphics[width=1.0\linewidth]{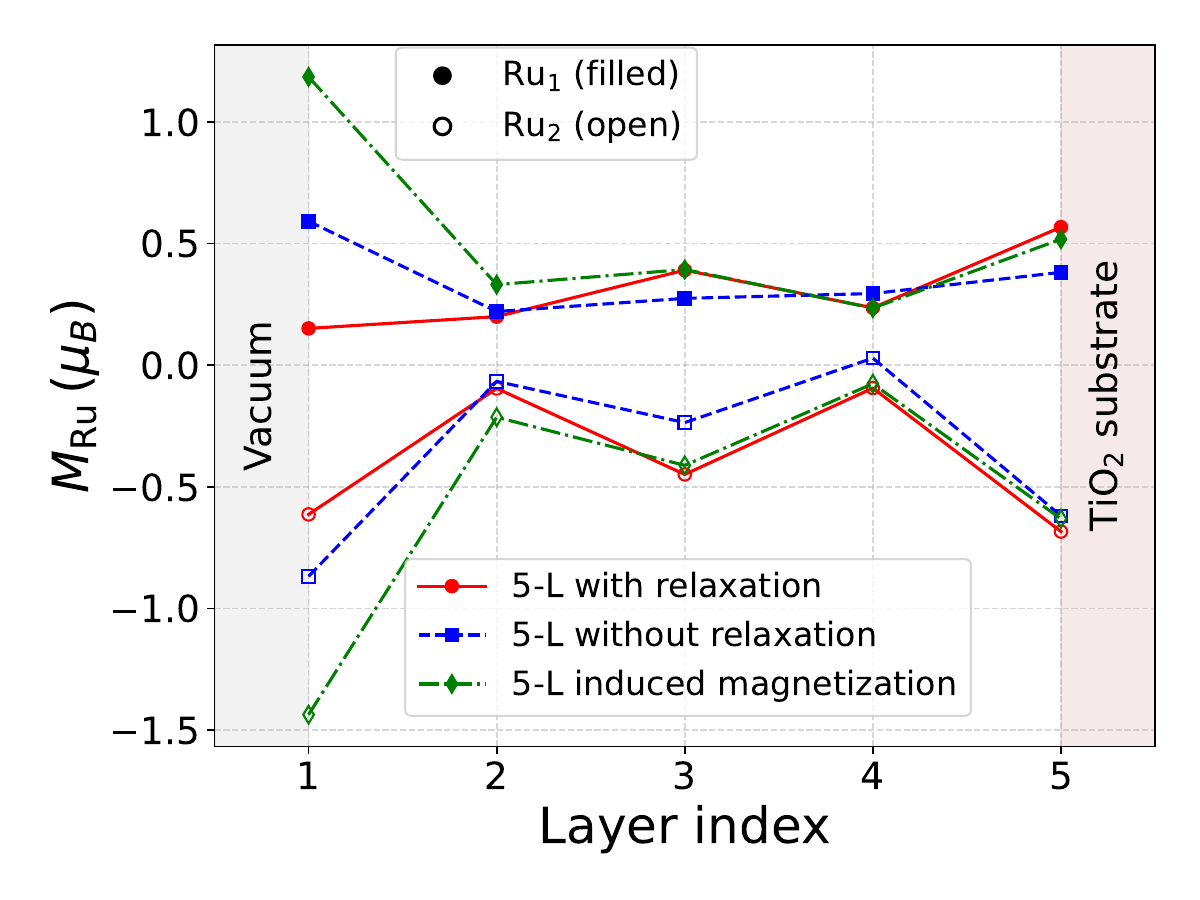}
    \caption{Effect of atomic relaxation and induced surface magnetization in a 5-L RuO$_2$(110) slab on TiO$_2$(110). Magnetic moments of Ru$_1$ (filled symbols) and Ru$_2$ (open symbols) are shown as a function of the Ru layer index. In the induced-magnetization scenario, we introduce a Hubbard $U$ correction on the surface Ru atoms to artificially enhance their magnetic moments. For this case, we use the geometrically relaxed structure.}
    \label{fig:m_ru110_5-L}
\end{figure}

Although full structural relaxations were carried out for all slab geometries, Fig.~\ref{fig:m_ru110_5-L} shows that atomic relaxation has only a minor effect on the magnetic moments of the inner RuO$_2$(110) layers. 
Moreover, when the magnetic moment of the surface layer is artificially enhanced by introducing a Hubbard $U$ correction on the Ru atoms at the surface, the induced increase remains strongly localized at the surface, while the magnetic moments in the inner layers change only negligibly. 
This behavior demonstrates that the magnetic moments generated at the surface due to symmetry breaking~\cite{sym-break2025} do not propagate into the inner layers. Instead, the magnetic moments of the Ru atoms are primarily determined by strain-induced changes in their local atomic environment arising from interfacial tensile strain, rather than by long-range induction from the surface.

\begin{figure}
    \centering
    \includegraphics[width=1.0\linewidth]{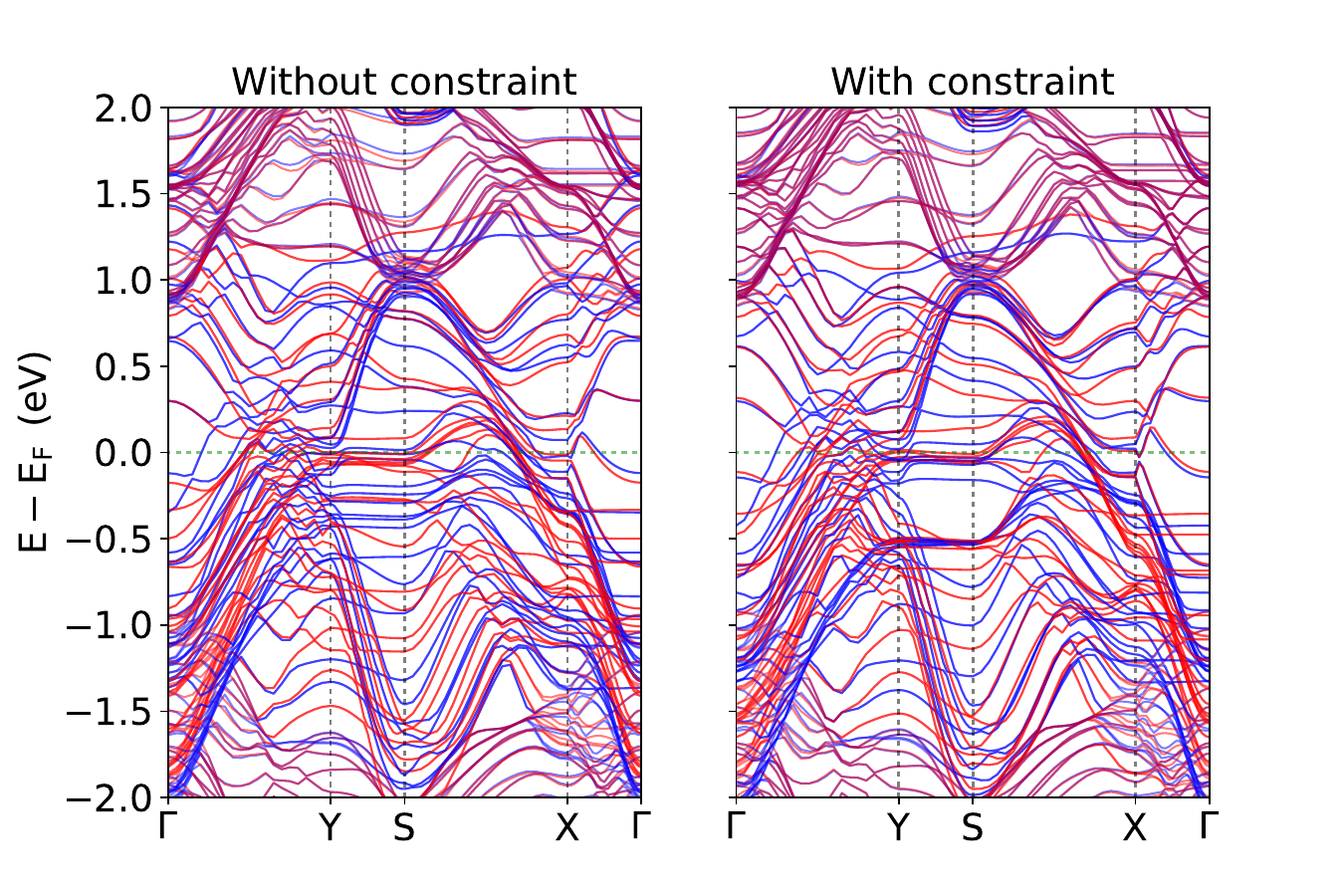}
    \caption{Nonrelativistic band structures of a 5-L RuO$_2$(110) slab on a TiO$_2$(110) substrate. The left panel shows the band structure of the unconstrained magnetic ground state (the true ground state with a ferrimagnetic-like ground state), while the right panel shows the band structure obtained with Ru magnetic moments constrained to enforce a fully compensated AFM configuration (a hypothetical altermagnetic ground state). In both cases, the non-relaxed atomic structure is used. Red and blue lines denote spin-up and spin-down bands, respectively.}
    \label{fig:band_struc}
\end{figure}

To further elucidate the magnetic character, we compare the electronic structures of a hypothetically compensated AFM (altermagnetic) RuO$_2$(110)/TiO$_2$(110) configuration and its true ferrimagnetic-like ground state. Band structures are calculated for a five-layer RuO$_2$(110) slab on TiO$_2$(110) in both magnetic configurations, using the same non-relaxed geometry to exclude structural relaxation effects; see Fig. \ref{fig:band_struc}.

To make the comparison between the two magnetic states meaningful, we adjusted the magnetic moments of the Ru atoms in the hypothetical altermagnetic case so that their total absolute magnetization matches that of the true magnetic ground state, where the magnetic moments vary from layer to layer (Fig.~\ref{fig:m_ru110}).  By testing different values, we found that fixing the Ru moments to approximately $0.3~\mu_\mathrm{B}$ reproduces the total absolute magnetization of the unconstrained ground state. This value is therefore used in the hypothetical altermagnetic state calculations presented in the right panel of Fig. \ref{fig:band_struc}.

As shown in Fig.~\ref{fig:band_struc}, the band structure of the true unconstrained ground state exhibits a substantially larger nonrelativistic spin splitting than that of the hypothetical altermagnetic state. This enhancement arises because, in the true magnetic ground state, the Ru magnetic moments are not fully compensated, in contrast to the compensated AFM limit, where the net magnetization vanishes.

\subsection{RuO$_2$(100)/TiO$_2$(100) and RuO$_2$(001)/TiO$_2$(001) heterostructures}
\begin{figure}
    \centering
    \includegraphics[width=1.0\linewidth]{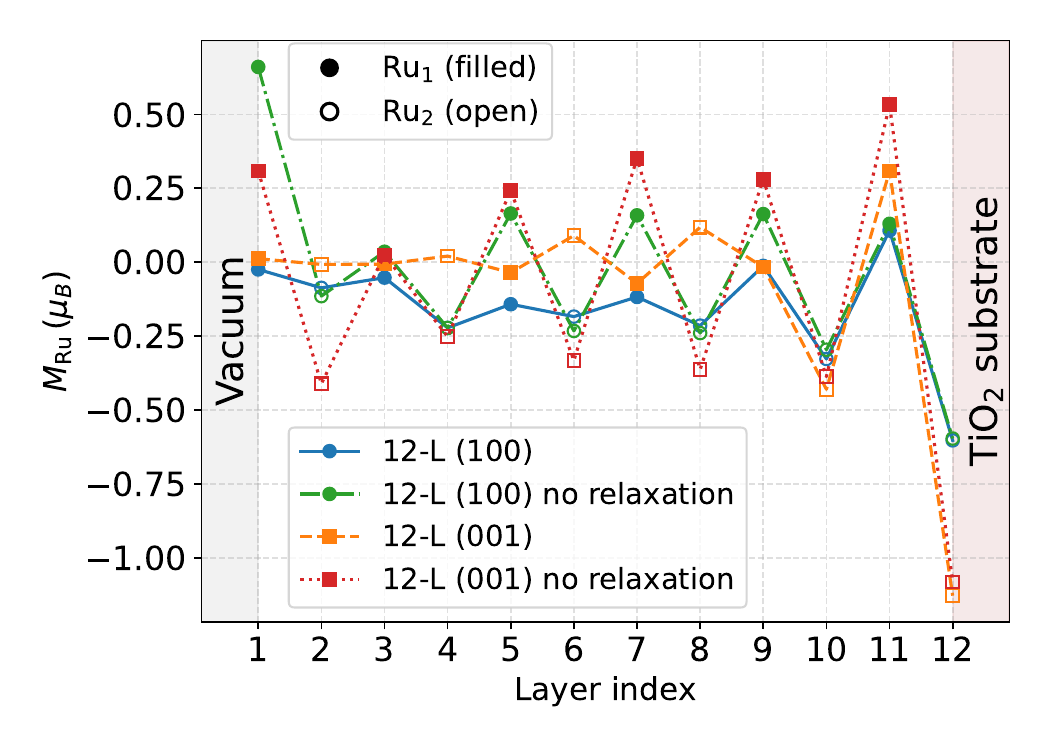}
    \caption{Layer-resolved Ru magnetic moments for a 12-layer RuO$_2$ slab with (100) and (001) orientations on a TiO$_2$ substrate. Filled and open symbols denote Ru$_1$ and Ru$_2$ sites, respectively. Results are shown for both relaxed and unrelaxed atomic structures.}
    \label{fig:m100-001}
\end{figure}

We next consider the remaining low-index RuO$_2$ surfaces, namely the (100) and (001) orientations. Owing to computational limitations, only one film thickness (12-L) is examined for each case. Figure~\ref{fig:m100-001} shows the layer-resolved magnetic moments of the two inequivalent Ru sites, Ru$_1$ and Ru$_2$, before and after structural relaxation.

For both orientations, atomic relaxation has a markedly stronger effect than in the (110) case. In the relaxed (001) and (100) films, magnetic moments in the outermost three layers are strongly suppressed, while inner layers near the RuO$_2$/TiO$_2$ interface develop  larger local moments. By contrast, the unrelaxed films exhibit an nearly compensated AFM order, with Ru$_1$ and Ru$_2$ moments of similar magnitude and opposite sign. 

Despite these differences, all systems, relaxed and unrelaxed, (001), (100), and (110), exhibit a characteristic layer-by-layer oscillation in the magnitude of the local magnetic moments. This behavior can be attributed to the metallic nature of the films: electronic states are confined between the surface and the interface, forming quasi-two-dimensional states analogous to a particle-in-a-box with finite potential barriers. The resulting quantum interference gives rise to oscillatory magnetic responses across the film thickness.

\subsection{Intrinsic magnetism of freestanding RuO$_2$ thin films}
To further explore the role of the substrate, we also study freestanding RuO$_2$ thin films with different surface orientations; see Fig.~\ref{fig:struct}. Two classes of films are considered: (i) films constrained to the experimental lattice parameters of bulk rutile TiO$_2$, and (ii) films constrained to the experimental lattice parameters of \emph{bulk} RuO$_2$. In both cases, only the internal atomic coordinates are relaxed, while the lattice parameters are fixed. 
Because the in-plane lattice constant of RuO$_2$ exceeds that of TiO$_2$ by approximately 0.1 Å, fixing RuO$_2$ to the TiO$_2$ lattice parameters introduces an effective biaxial tensile strain. This procedure, therefore, serves as a computational strategy to simulate RuO$_2$ under tensile strain conditions.

\begin{figure}
    \centering
    \includegraphics[width=1.0\linewidth]{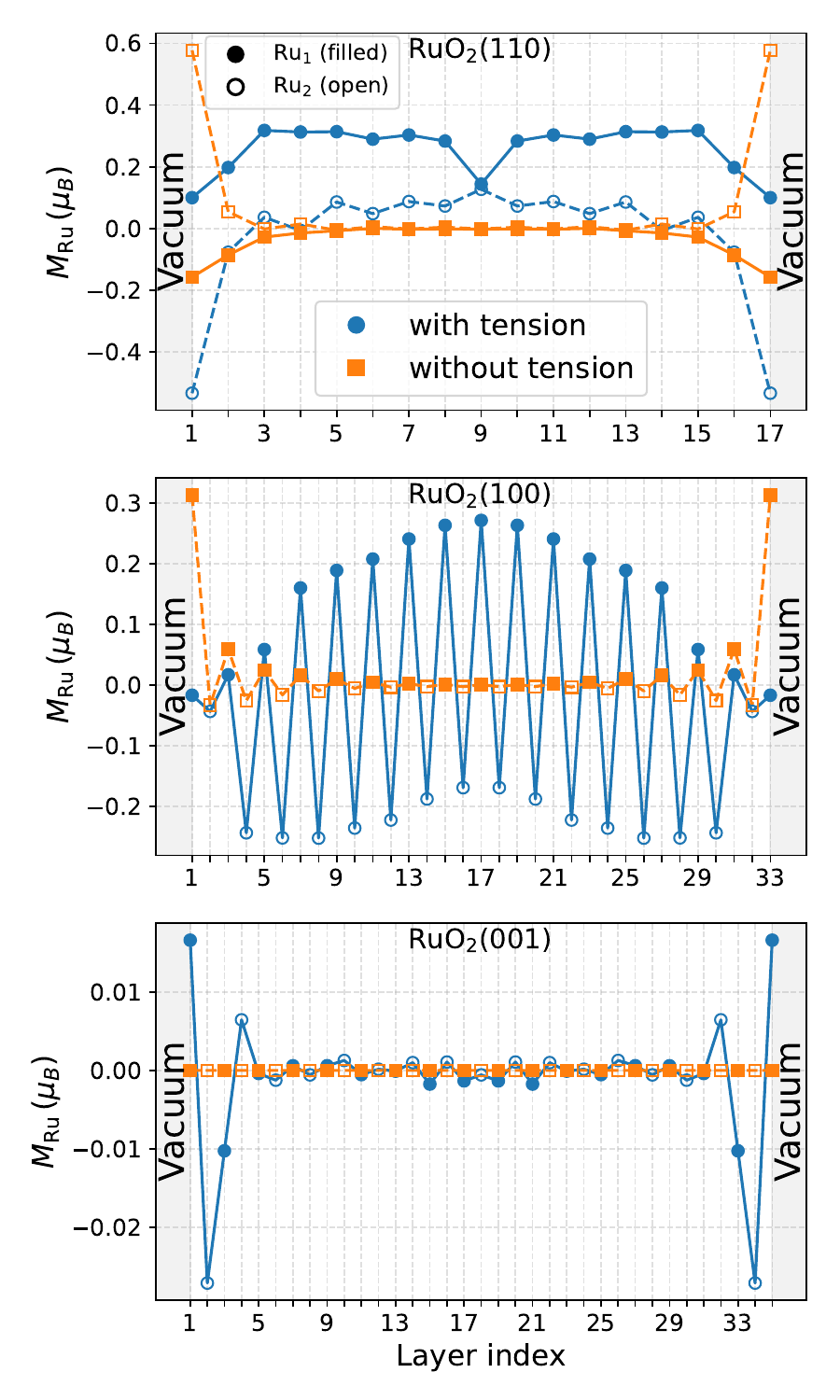}
    \caption{Layer-resolved Ru magnetic moments for freestanding RuO$_2$ slabs with (110), (100), and (001) orientations. For each orientation, two unrelaxed reference structures are shown: one constrained to the experimental TiO$_2$ lattice parameters (tensile strain) and one using the experimental bulk RuO$_2$ lattice. Filled and open symbols denote Ru$_1$ and Ru$_2$ sites, respectively.}
    \label{fig:RuO2-only}
\end{figure}

Figure~\ref{fig:RuO2-only} shows the layer-resolved Ru magnetic moments for the (110), (100), and (001) orientations, both with and without tensile strain. 

In the unstrained (110) and (100) films, finite magnetic moments are largely confined to the first two surface layers and decay rapidly toward the inner layers. The unstrained (110) result is consistent with a recent report Ref.~\cite{sym-break2025}.
In contrast, the (001) surface converges to a nonmagnetic solution, with no Ru atom exhibiting a discernible magnetic moment within all layers. This finding differs from a recent report of finite magnetic moments at the (001) termination~\cite{magnetization_001}. 

To evaluate the robustness of our results in the unstrained case, we carried out additional calculations using VASP with several different PBE pseudopotentials for Ru, namely Ru, Ru\_pv, and Ru\_sv, corresponding to the valence configurations $4d^7 5s^1$, $4p^6 4d^7 5s^1$, and $4s^2 4p^6 4d^7 5s^1$, respectively, as specified on the VASP website. All three yield the same outcome (unshown); for the unstrained (001) surface, the magnetic moments remain negligible, $<10^{-6}\mu_{\mathrm{B}}$ per Ru atom. We therefore attribute the discrepancy with Ref. \cite{magnetization_001} to methodological differences, most likely related to the Brillouin-zone integration scheme, as our analysis indicates that the emergence of magnetism in RuO$_2$ is highly sensitive to the details of the $k$-space sampling.

Under tensile strain, finite magnetic moments develop for all three surface orientations. However, the response of the (001) surface is markedly weaker: Ru atoms in the top layer carry moments of only $\sim$0.02–0.3$\mu_{\mathrm{B}}$, which decay rapidly in the subsurface layers. 

For RuO$_2$(110) thin films, tensile strain qualitatively alters the magnetic response relative to the unstrained case. In the topmost layers, the magnetic moments of the inequivalent Ru sites (Ru$_1$ and Ru$_2$) acquire opposite signs. However, deeper in the slab, the Ru$1$ moment remains nearly constant at $\sim$0.3$\mu{\mathrm{B}}$, whereas the Ru$2$ moment reverses sign and becomes small and positive $<0.1,\mu{\mathrm{B}}$. This pronounced layer- and site-dependent evolution underscores the complex strain-induced magnetism of the (110) surface.

For RuO$2$(100) thin films, tensile strain induces the strongest tendency toward AFM alignment among the orientations studied; however, the system remains far from a fully compensated state. Our DFT calculations yield a sizable net magnetic moment of approximately 4.63$\mu{\mathrm{B}}$ per unit cell.

These results are in contrast to a recent report in Ref. \cite{Strain_Engineering} that found the (100) orientation to be an ideal altermagnet and the (110) orientation to be ferrimagnetic. Based on our accurate analysis, none of the investigated surfaces, (110), (100), or (001), stabilize a compensated AFM and hence an altermagnetic ground state. Again, we attribute this discrepancy to the methodology used in their \emph{ab initio} calculations. 

\section{summary and conclusion}
In this work, we have performed a comprehensive first-principles investigation of the magnetic properties of RuO$_2$ thin films with (110), (100), and (001) orientations, both as RuO$_2$/TiO$_2$ heterostructures and (un)strained freestanding slabs. Motivated by the ongoing debate on the existence of $d$-wave altermagnetism in metallic RuO$_2$, we systematically examined the role of epitaxial strain, surface termination, film thickness, structural relaxation, and Brillouin-zone integration schemes in determining the magnetic ground state.

Our calculations demonstrate that the true magnetic ground state in RuO$_2$ thin films is intrinsically fragile and highly sensitive to both structural and methodological details. In particular, we show that the calculated magnetic moments depend strongly on the choice of the Brillouin-zone integration scheme and smearing parameters. When treated with sufficiently accurate integration, bulk RuO$_2$ at its experimental lattice parameters remains nonmagnetic, consistent with several recent experimental reports that question robust long-range antiferromagnetism in this compound. This sensitivity places RuO$_2$ in close proximity to a magnetic instability, where small perturbations can induce local moments.

In RuO$_2$(110)/TiO$_2$(110) heterostructures, the magnetization is dominated by surface and interface layers and exhibits pronounced oscillations with film thickness, consistent with quantum confinement and modified Ru–O hybridization at the boundaries. Even when initialized in various magnetic configurations, the system relaxes to states characterized by residual net magnetization, i.e., an unusual ferrimagnetic-like order rather than a compensated AFM order with altermagnetic behavior.

For freestanding and unstrained (110) and (100) slabs, only weak surface-localized moments develop, while the unstrained (001) surface remains essentially nonmagnetic within numerical accuracy. 

Tensile biaxial strain, mimicked by constraining freestanding RuO$_2$ thin film to the rutile TiO$_2$ lattice parameters, induces finite Ru magnetic moments for all investigated surface orientations. However, in no case do we find stabilization of a compensated Néel-ordered AFM order and hence any altermagnetic behavior. The (100) orientation exhibits the largest net magnetic moment under strain, while the (001) orientation shows the weakest magnetic response overall.

Through a direct comparison between a hypothetical compensated AFM configuration, i.e., altermagnet state, and the true magnetic ground state, i.e., ferrimagnetic-like state, we show that the pronounced spin splitting in thin films stems from the uncompensated magnetization, not from symmetry-enforced altermagnetism. Therefore, experimentally observed spin splitting or anomalous transport in RuO$_2$ thin films should not be uncritically attributed to $d$-wave altermagnetism, but instead evaluated in view of possible emerging unconventional ferrimagnetic-like states in the system.

Overall, our results reconcile seemingly conflicting various theoretical and experimental findings by demonstrating that RuO$_2$ resides at the brink of magnetism: methodological differences, small strain, or reduced dimensionality, can induce small local magnetic moments; yet none of the studied realistic thin-film geometries stabilize a compensated altermagnetic ground state. These findings underscore the importance of rigorous numerical convergence and detailed structural modeling when identifying candidate altermagnets, and they suggest that achieving robust altermagnetism in RuO$_2$ may require additional symmetry engineering, chemical modification, or external tuning beyond simple epitaxial strain.

In addition, our findings call for a careful re-evaluation of recent anomalous Hall measurements in RuO$_2$ thin films that have been interpreted within a fully compensated altermagnetic framework. Since our calculations reveal an unconventional ferrimagnetic-like state with finite net magnetization in epitaxial RuO$_2$ thin films, implying conventional Berry-curvature-driven anomalous Hall effects. Given the strong spin–orbit coupling in RuO$_2$/TiO$_2$ heterostructures, we propose systematic thickness- and strain-dependent transport measurements to distinguish genuine symmetry-enforced altermagnetism from the uncompensated magnetism predicted here.

\section{Method}
We performed nonrelativistic, \emph{{ab initio} } density functional theory (DFT) calculations using the Quantum ESPRESSO (QE) package~\cite{QE-2009,QE-2017,QE-2020}. The electron–ion interactions were described using the Standard Solid-State Pseudopotentials (SSSP) library, PBEsol Efficiency version 1.3.0~\cite{SSSP}.
For the exchange–correlation functional, we employed the PBEsol approximation~\cite{PBEsol}, a revised form of the Perdew–Burke–Ernzerhof (PBE) generalized gradient approximation (GGA) that provides improved accuracy for the structural properties of solids and surfaces. A vacuum spacing of approximately 15~\AA{} was introduced in all thin-film calculations to avoid spurious interactions between periodic images.
Brillouin-zone integrations were performed using a uniform $k$-point grid corresponding to a reciprocal-space resolution of 0.10~\AA$^{-1}$. The thin-film structures were constructed using the CIF2Cell code~\cite{cif2cell}, and the initial QE input files were generated with the QE input generator~\cite{SSSP}.
The plane-wave kinetic energy cutoff was set to 40~Ry for the wave functions and 400~Ry for the charge density. 
The magnetic moment associated with each Ru atom was obtained using Bader charge analysis as implemented in the Critic2 code~\cite{Critic_2009, Critic2_2014}. 
\\
\section*{Data availability}
The data that support the findings of this study are available upon reasonable request.
\section*{Author Contributions}
M.A. and A.Q. jointly defined the project. M.A. conceived and supervised the research, performed the majority of the DFT calculations, and co-wrote the manuscript. N.R. carried out structural visualizations, contributed to selected DFT calculations, and provided intellectual input. I.M. performed additional simulations. A.R.O. provided intellectual input and financial support. M.A. and A.Q. jointly discussed the results and finalized the manuscript. All authors reviewed and edited the manuscript.

\section*{Competing interests}
The authors declare no competing interests

\section*{ACKNOWLEDGMENTS}
A.Q. has been supported by the Research Council of Norway through its Centers of Excellence funding scheme, Project No. 262633 “QuSpin” and FRIPRO with Project No. 353919 “QTransMag.”

\bibliography{ref.bib} 

\end{document}